# Comparative performance of three optical biosensing platforms for SARS-CoV-2 antibodies detection in human serum

*Agostino Occhicone[1,2,‡], Alberto Sinibaldi[1,2,‡,*], Peter Munzert[3], Jordan N. Butt[4], Ethan P. Luta[6], Diego M. Arévalo[4], and Francesco Michelotti[1], and Benjamin L. Miller[5,6]*

1. SAPIENZA Università di Roma, Dipartimento di Scienze di Base e Applicate per l'Ingegneria, Via A. Scarpa 16 00161, Roma, Italy

2. Italian Institute of Technology, Centre for Life Nano and Neuro Science, Viale Regina Margherita 291, 00161, Roma, Italy

3. Fraunhofer Institute for Applied Optics and Precision Engineering IOF, Albert-Einstein-Str. 7, Jena 07745, Germany

4. Department of Chemistry, University of Rochester, Rochester, NY 14627, USA

5. Department of Dermatology, University of Rochester Medical Center, Rochester, NY 14642, USA

6. Materials Science Program, University of Rochester, Rochester, NY 14642, USA

---

* Author to whom correspondence should be addressed to francesco.michelotti@uniroma1.it

**Abstract**

This study presents a rigorous comparative analysis of two label-free optical biosensing platforms, Bloch surface wave (BSW) and microring resonator (MRR), for the detection of SARS-CoV-2 antibodies in human serum. To ensure direct comparability, a new BSW readout system was established alongside an existing MRR platform, allowing assays to be conducted under nearly identical experimental conditions. Both sensors were functionalized with various SARS-CoV-2 Spike and Nucleocapsid protein variants to capture specific host antibodies. The results demonstrate that both platforms provide rapid, quantitative, and sensitive detection of anti-Spike and anti-Nucleocapsid antibodies without the need for secondary labels. Furthermore, the platforms show excellent agreement with longitudinal serology benchmarks and high repeatability across different biochip batches. This work establishes both BSW and MRR technologies as powerful, low-cost candidates for next-generation clinical diagnostics and serological surveillance.

**INTRODUCTION**

The emergence of the Severe Acute Respiratory Syndrome Coronavirus 2 (SARS-CoV-2) in 2019 underscored the critical need for advanced diagnostic and serological tools.[1] While reverse transcription-polymerase chain reaction (RT-PCR) tests remain the gold standard for detecting active viral infection[2], serological assays that quantify host antibodies are indispensable for assessing immune responses, determining vaccine efficacy, tracking herd immunity, and understanding the epidemiology of a viral infection[3–5].

In response to these challenges, the field of optical biosensing has offered powerful, real-time, and label-free solutions for biomolecular interaction analysis. Among these, both Bloch surface wave[6] (BSW) and microring resonator[7] (MRR) biosensors have been demonstrated as highly efficient, sensitive analytical tools for SARS-CoV-2 antibody detection in human serum. Both technologies can be integrated into disposable consumable formats with a low cost of production.

BSW biosensors represent a promising frontier in label-free and fluorescence optical detection.[8–16] Unlike surface plasmon polaritons (SPP) sustained by metal layers,[17,18] BSWs can propagate along the surface of a purely dielectric and low loss one-dimensional photonic crystal (1DPC). This fundamental difference results in exceptionally sharp optical resonances, corresponding to a label-free resolution that is comparable to conventional surface plasmon resonance (SPR) platforms.[19] Moreover, the lack of quenching due to metals favors their application in fluorescence-based detection schemes.[20–26]

MRR similarly constitutes a highly useful approach to biological detection. In these devices, light evanescently couples from a linear bus waveguide into a circular waveguide, where it gains phase. Binding of a target analyte to an immobilized capture molecule on the surface of the ring changes the effective refractive index, resulting in a shift of the resonant wavelength. Widely studied for many years[7,27–32], MRR have been shown to be a viable approach for label-free biosensing. We have used MRRs for multiplex detection of antibodies to SARS-CoV-2 and influenza[7], described an MRR approach to simultaneous measurement of enzymatic cleavage and antibody binding[33], and recently demonstrated an MRR approach for rapid diagnosis of Von Willebrand disease[34], all while being referenced to a non-specific control ring[35].

In this work, we present the results of a comparative study of the performance of BSW and MRR biosensors for the direct and label-free detection of SARS-CoV-2 anti-Spike and anti-Nucleocapsid protein antibodies in human serum samples. For this purpose, a new apparatus for BSW-based label-free detection was set up at the University of Rochester Medical Center on the same optical table where a MRR platform is routinely operating. The objective was to perform comparative assays under as identical conditions as possible, thereby eliminating heterogeneities arising from experiments conducted in different laboratories and from the transfer of samples between them.

In all cases, we functionalized the biosensors' surface either with the Receptor Binding Domain (RBD) of different variants of the SARS-CoV-2 Spike protein (wild type, Omicron, BA-5) or with the wild type Nucleocapsid protein, to ensure specific capture of relevant antibodies. The platforms' performance was rigorously evaluated via assays that were performed with the same serum samples and with the same handling procedures by different operators. For one specific assay the results were also benchmarked against those obtained with a commercial SPR platform. While we do observe differences in performance characteristics (likely a result at least in part of differences in their fluidic interfaces), both the BSW and MRR platforms demonstrated their capability for rapid and quantitative antibody detection without the need for secondary labels. As such, this study highlights the distinct advantages of both the BSW and the MRR platforms and establishes their significant potential as next-generation tools for clinical diagnostics and sero-epidemiological surveillance.

## MATERIALS AND METHODS

**Serum Samples.** For serum collection, whole blood was drawn via venipuncture, allowed to clot at ambient temperature for at least 1 hour, and then centrifuged at 1200 x g for 15 min. Serum was drawn off via pipette, aliquoted, and stored at -80 ˚C prior to use. Sera were drawn under protocols approved by the University of Rochester Medical Center Institutional Review Board for Dermatology Department assay development. Written consent was obtained from all subjects. For the present study we focused on three specific samples from a larger cohort. Two samples were selected based on results of a previous study,[36] in which a fully automated commercial Arrayed Imaging Reflectometry system (ZIVA, by Adarza Biosystems, Inc) was used to perform reliable longitudinal serology against a 34-plex respiratory array, including SARS-CoV-2. In Table 1 we list the serum samples used here with the respective donor history and ZIVA scores, where available.

*Table 1 – Characteristics of serum samples used in the present study.*

| Code | Patient history | ZIVA values | |
|---|---|---|---|
| | | N-wt | S-wt |
| S404 | Vaccinated mRNA (2 doses of PFIZER in Jan 2021). SARS-Cov-2 infection (PCR positive) in Nov 2021 (likely Delta infection). Sample drawn on Dec 21$^{st}$ 2021, 1 month post recovery | 0.8 ± 0.4 | 40 ± 1 |
| S405 | Infected (Apr 2020, PCR positive). Vaccinated (2x Pfizer doses mRNA Jan 2021). Infected again Jan 2022 (likely Omicron infection). Sample drawn: Dec 1$^{st}$ 2022, 1 week post recovery | 31 ± 1 | 22.3 ± 0.7 |

**BSW biochips and readout principle.** The BSW biochips consist of dielectric multilayers deposited on standard microscope slides by plasma ion assisted evaporation under high vacuum conditions and were described elsewhere.[37] The layers are made of $SiO_2$, $Ta_2O_5$, and $TiO_2$. Starting from the substrate the stack sequence is $SiO_2/(Ta_2O_5/SiO_2)^2/TiO_2/SiO_2$, with thicknesses, expressed in nm, 275/(120/275)²/20/20. At $\lambda = 632.8\ nm$, the complex refractive indices are $n_{SiO_2} = 1.447 + i5 \times 10^{-6}$, $n_{Ta_2O_5} = 2.096 + i5 \times 10^{-5}$ and $n_{TiO_2} = 2.292 + i1.8 \times 10^{-3}$.

To operate under total internal reflection (TIR) conditions in the Kretschmann configuration,[38] the back of the biochip substrate was coupled to a BK7 glass prism ($n_p$ = 1.515) using a contact oil, as shown in the inset of Figure 1. The optical configuration of the readout platform, that was purposely setup for the present work, derives from previous work[39] and is described in detail in the Supplementary Information (SI). A TE-polarized laser beam at λ = 632.8 nm is focused by a cylindrical lens into a line at the surface of the BSW biochip, probing the reflectance with spatial resolution along a microfluidic channel filled with an aqueous medium. The reflected beam is collected by a cylindrical Fourier lens and imaged onto a CMOS array detector, thus providing an angularly resolved resonance detection scheme[40]. An example image from the CMOS camera is shown in the inset of Figure 1(a). The 64 rows represent different positions along a 6.8 mm-long line, with the excitation of a BSW resonance observed as a dip in the reflectance profile along the row.

A microfluidic channel was defined on top of the BSW biochips using a PDMS sheet that was cut to define a 4 mm wide and 130 μm thick channel and then topped with a PDMS block with tube fittings. Bonding of the PDMS sheet to the biochip surface was favored by a plasma-ozone pretreatment. All analyte injections were performed at 12.5 μl min⁻¹, accommodating reaction volumes of 120–160 µl, whereas washing and regeneration were conducted at 20 µl min$^{-1}$ with volumes up to 250 µl.

**MRR biochips and readout platform.** The MRR were fabricated as previously described[7], and consist of a 5 µm SiO$_2$ lower cladding, and a waveguide core comprised of silicon nitride that is 220 nm thick and 1.5 µm wide designed for TE-mode operation. Eight rings were fit within an 800 µm wide surface, with 165 µm diameter rings and 375 nm coupling gaps for the rings themselves. The optical setup and the injection molded micropillar cards are described in detail in the SI. All MRR assays incorporated a 15 µl pre-wash (Assay Wash Buffer, AWB; defined below), to hydrate the chip and wash away dried salts and stabilizer, followed by 30 µl of analyte in a passive microfluidic process.

**SPR biochips and readout platform.** The benchmark assays with SPP based sensors were carried out with a Cytiva Biacore X100 SPR instrument. Ready-to-use SPR biochips were purchased from Cytiva (sensor chip CM5) and featured a standard dextran surface chemical functionalization. A pH scouting study was conducted to optimize the immobilization of RBD protein on the surface of the SPR chip. A pH of 5 was determined to be the optimal pH. Proteins (50 µg ml$^{-1}$ in pH 5 sodium acetate buffer) were then immobilized using the common 0.4 M EDC (1-ethyl-3-(3-dimethylaminopropyl)carbodiimide) and 0.1 M NHS (N-hydroxysuccinimide) immobilization chemistry on flow cell 2. Flow cell 1 was blocked with 1 M ethanolamine and used to account for non-specific binding interactions with the CM5 surface[41].

**Molecular probes and standards.** The following recombinantly expressed (baculovirus) SARS-CoV-2 antigens were purchased from Sino Biological, Inc. (Wayne, PA): Wild-type Spike RBD (His Tag) recombinant protein (S-wt, Cat: 40592-V08H), Omicron Spike RBD protein (His Tag) (S-O, Cat: 40592-V08H121), BA.5 Spike RBD Protein (His Tag) (S-BA.5, Cat: 40592-V08H130),

Nucleocapsid (His Tag) recombinant protein (N-wt, Cat: 40588-V08B). SARS-CoV-2 Omicron (BA.4/BA.5/BA.2.75) Spike RBD Specific Antibody, Chimeric Mab (a-BA.5, Cat: 40589-D003) was purchased from Sino Biological, Inc. too. Antifluorescein (a-FITC) antibody used as a nonspecific binding control was obtained from Rockland Immunochemicals (Limerick, PA). The diluent for antibody/antigen printing was modified (potassium-free) phosphate-buffered saline (mPBS, pH 7.2). Assay wash buffer (AWB), which was used to dilute serum samples, consisted of mPBS with 3 mM EDTA and 0.01% Tween-20. StabilGuard Immunoassay Stabilizer was obtained from Surmodics IVD Inc., Eden Prairie, MN. All serum samples were diluted 1:5 or higher, as noted, in AWB.

**Chemical functionalization and bioconjugation of the BSW and MRR biochips.** BSW and MRR biochips were first washed for 15 min in a 1:1 mixture of methanol and concentrated hydrochloric acid, then washed for 15 min in 3:1 mixture of concentrated sulfuric acid and 25% hydrogen peroxide, then rinsed 5 × 30 s in Nanopure water and dried with nitrogen. The biochips were next placed in a chemical vapor deposition (CVD) system (Yield Engineering Systems, Fremont, CA), where a monolayer of (3-Glycidyloxypropyl)trimethoxysilane (GPTMS; Gelest, Inc., Morrisville, PA) was deposited on the surface. Ellipsometry was used to confirm the deposition of GPTMS on sentinel Si/SiO$_2$ chips; thickness measurements typically indicated a deposited film of 5-10 Å.

Antigens and control antibodies were covalently attached to the functionalized surface by spotting them directly on the sensing regions using a sciFLEXARRAYER SX piezoelectric microarray (Scienion AG, Berlin, Germany) using the manufacturer's Find Target Reference Points (FTRP) machine vision protocol to accurately locate the position onto the biochips' surface, which is particularly important for MRR biochips. Capture proteins were buffer exchanged and stored in mPBS (pH 7.2) and printed at a concentration of 0.4 mg ml$^{-1}$, and anti-fluorescein at 0.55 mg ml$^{-1}$.

**Regeneration solution.** The BSW biochips used in the assays shown in Figure 3 and Figure 4 were regenerated at a flow rate of 20 µl min$^{-1}$ using a 10 mM aqueous solution of Glycine•HCl and citric acid to achieve pH 1.8.

**RESULTS AND DISCUSSION**

We first discuss results of assays carried out with BSW biochips. These results were obtained by means of a BSW read-out platform that was purposely set up at the University of Rochester to enable direct comparison with the MRR platform. Then we shall compare the results of assays carried out with either BSW or MRR biochips with the same human serum. Finally, we shall discuss a comparison between BSW, MRR and SPR results for a specific assay.

When comparing the results of assays carried out with these three biosensors, especially the binding kinetics curves, one should keep in mind that the microfluidics used in the comparative assays are different due to the specific biosensors' configurations. Moreover, the chemical surface functionalization strategies used in the BSW/MRR cases and the SPR case are different, involving capture layers characterized by a different thickness.

**Assays with BSW biochips**

Results of an assay carried out with the 1:5 diluted S404 serum are shown in Figure 1. Before topping the BSW biochip with the microfluidic cell we immobilized the proteins S-wt, N-wt (signal regions) and anti-FITC (negative control region) in three different regions, as sketched in the inset of Figure 1(b). In the CMOS image of the reflected beam shown in the inset of Figure 1(a), the BSW resonance is shifted to different angular positions in the three sensitive regions, due to the different density of immobilized probes.

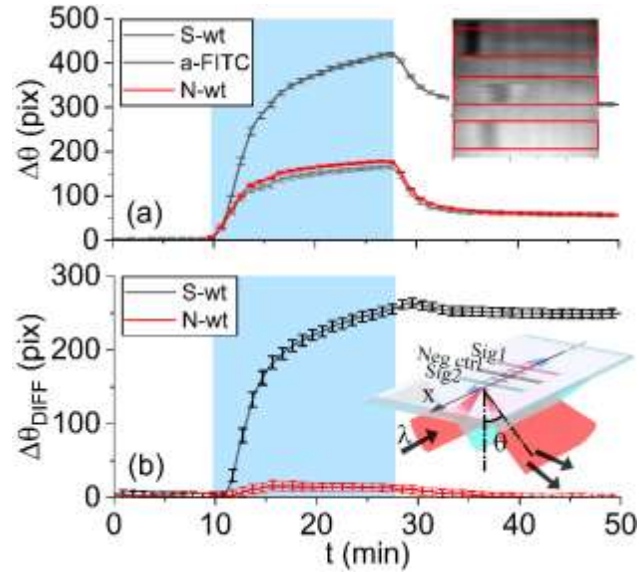

*Figure 1. Sensograms recorded during a label-free assay with a BSW biochip and serum S404. (a) Shift of the BSW resonance Δθ with respect to the level recorded in buffer before the serum injection, which is highlighted in blue. The sensograms correspond to the three regions where S-wt, a-FITC and N-wt were immobilized. Inset: CMOS image recorded at the beginning of the assay in buffer; the BSW resonance is shifted in different sensitive regions. The error bars are the standard deviation of the mean over the CMOS rows inside a region (b) Differential sensograms Δθ$_{DIFF}$ obtained by subtracting the signal recorded in the a-FITC control region from the two signal regions, S-wt and N-wt. Inset: Sketch of the optical read-out scheme.*

Figure 1(a) shows the time dependency of Δθ(t), which is the relative shift of the angular position of the BSW resonance with respect to the stable starting level recorded at the beginning of the assay when the biochip is filled with the AWB. The blue shaded area corresponds to the injection phase of the S404 serum, followed by the injection of the AWB for washing. It is clear that during the sample injection and contemporary incubation the BSW resonance shifts in all three regions as a result of the change of the bulk refractive index of the sample and of the change of the refractive index at the biochip surface, due to both specific and nonspecific binding of serum proteins. After washing, a residual shift is observed. The presence of nonspecific binding in the control region confirms what was previously observed in similar experiments with high-protein sera.[6] The signal recorded in the N-wt region is almost the same as that observed for the negative control, while that recorded in the S-wt region is much larger. Figure 1(b) shows the differential sensograms $\Delta\theta_{DIFF}$ obtained by subtracting the signal recorded in the negative control region from those recorded in the signal ones. Clearly the S-wt region responds in a much stronger way with respect to the N-wt region, consistent with the characteristics of serum S404, and the corresponding ZIVA values listed in Table 1, which was drawn from a subject who received a mRNA vaccine (high S-wt content) but had never been infected by the virus (expected low N-wt content). The residual differential signals are stable after washing, as expected for a reliable assay.

Figure 2 shows the differential sensograms $\Delta\theta_{DIFF}$ recorded in assays carried out with three different pristine BSW biochips and with the same 1:5 diluted S405 serum sample. In all biochips the negative control is a region with a-FITC. In the signal regions of the three BSW biochips we immobilized different couples of SARS-CoV-2 related proteins: (a) S-wt, N-wt; (b) N-wt, S-O; (c) S-BA.5, S-O. The comparative analysis of the three assays provides the following observations. In contrast to the data for S404 shown in Figure 1, results for S405 shown in Figure 2(a) indicate that the response of the N-wt region is stronger than the S-wt region. This result is in agreement with the history of the S405 serum sample and the ZIVA values reported in Table 1: in this case infection preceding vaccination led the immune system to generate a large amount of anti-Nucleocapsid antibodies in response to the donor's exposure to the complete virus. Comparison of Figure 2(a) and (b) highlights that the $\Delta\theta_{DIFF}$ of the N-wt region of two different BSW biochips (red curves), prepared and used on different days with the same procedures, is the same within the limit of the experimental error, underscoring the batch to batch repeatability of the BSW biochips. Comparison of Figure 2(b) and (c) confirms the repeatability of the results obtained with different biochips with the same S-O probe region (blue curves). Overall, the three assays show that the human antibodies towards the S protein bind with comparable efficiency to the Spike variants S-wt, S-O and S-BA.5.

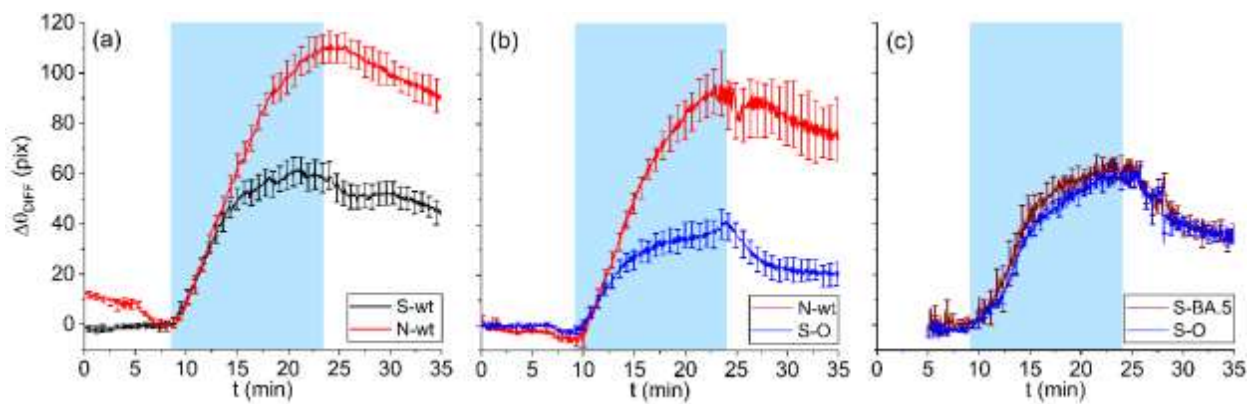

*Figure 2. Differential sensograms $\Delta\theta_{DIFF}$ recorded in assays carried out with three different pristine BSW biochips on different days with the same serum S405. The biochips were prepared by immobilizing different probes in the two signal regions: (a) S-wt and N-wt; (b) N-wt and S-O; (c) S-BA.5 and S-O. In all three cases we immobilized a-FITC in the negative control region.*

Regeneration of BSW biochips was also investigated, in order to assess whether they could be re-used in multiple antibody detection assays. Figure 3 shows the differential sensograms recorded during a label-free assay carried out with a single BSW biochip and the S405 serum. On the BSW biochip we immobilized S-wt, N-wt and a-FITC as a negative reference in three different regions. The assay started in the AWB, then we performed repetitively the following injection protocol: diluted serum sample (light blue)/AWB/regeneration solution (light brown)/AWB. Upon repeating the protocol, we used S405 serum samples with increasing dilution ratio (increasing serum concentration), from 1:1000 to 1:5.

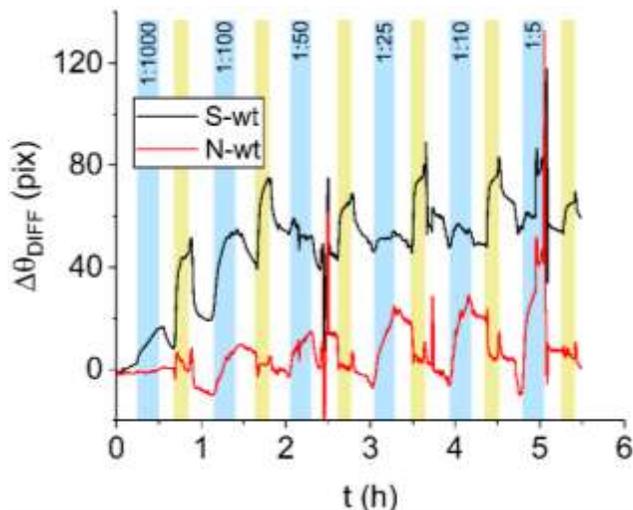

*Figure 3. Serial differential sensograms $\Delta\theta_{DIFF}$ recorded in an assay carried out with the same BSW biochip by injecting S405 diluted sera with increasing dilution ratio, from 1:1000 to 1:5. The biochip was prepared by immobilizing either S-wt or N-wt in the two signal regions and a-FITC in the negative control region. The injection of the sera and of the regeneration solution is marked with either a light blue box or a light brown box, respectively.*

Upon injecting the highly diluted sera (1:1000, 1:100) the label-free response is larger in the S-wt region than in the N-wt region, indicating a larger proportion of serum antibodies for S-wt. However, after the injection and incubation of the regeneration solution and washing in AWB there is no recovery of the differential signal to the start value in the S-wt region, while there is a recovery in the N-wt region. This last result shows that the regeneration solution cannot lead to unbinding of antibodies from the S-wt regions, probably owing to their larger affinity for S-wt. Moreover, during the injection and incubation of the regeneration solution the sensograms of the S-wt and N-wt regions differ substantially; while in the S-wt we observe a signal increase due to the change of the bulk refractive index of the solution, in the N-wt such an increase is missing, probably due to the contemporary decrease of the refractive index at the biochip surface due to unbinding of the antibodies. For larger dilution ratios (from 1:50 to 1:5) of the S405 serum, it appears that the S-wt region reached saturation of the capture probe sites, which are not regenerated, and that the response is missing upon injecting the diluted S405 sera. On the other hand, the N-wt region recovers the background level after each regeneration and shows an increasing signal upon injection and incubation of the sample. As a conclusion, we can infer that the BSW biochips cannot be re-used when S-wt probes are incubated at their surface, whereas they could be re-used in the case of the N-wt regions, provided an accurate characterization of the residual sensitivity is carried out after the regeneration step.

As far as the dependency of the BSW biochips response on the dilution of sera is concerned, we performed several different assays with different biochips that were prepared under the same conditions by immobilizing in the signal regions either S-wt or N-wt. As usual a-FITC was used as a reference. The assays were performed with the S405 serum, which was diluted from 1:100 to 1:5. While the assays with dilution 1:5 and 1:10 were performed with different pristine BSW biochips, the assays with dilution 1:25, 1:50 and 1:100 where performed with the same BSW biochip in a series and regenerating the surface with the same procedure shown in Figure 3. Therefore, based on the results on regeneration of the N-wt and S-wt regions reported above, the sensograms are acceptably reliable in the case of N-wt while they could be affected by a larger error in the case of S-wt. However, since during this specific assay (dilution up to 1:25) we did not reach the protein concentration of either crude or slightly diluted sera, they can be at least inspected to derive further information. In Figure 4(a) and (b), we show the sensograms recorded for different dilutions of the S405 serum, in either the N-wt or the S-wt region, respectively. The results confirm that, for the S405 serum, the response of the N-wt region is larger than in the S-wt region. The sensograms' amplitude increases for larger dilution ratios (higher serum content), as expected.

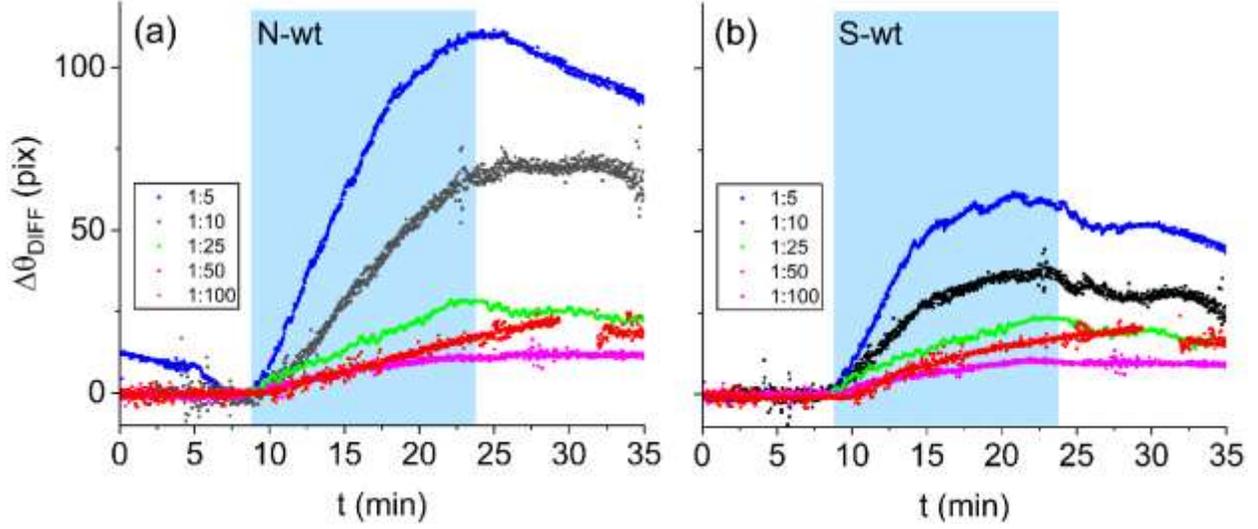

*Figure 4. Differential sensograms Δθ$_{DIFF}$ recorded in assays carried out with different BSW biochips when injecting S405 diluted sera with different dilution ratios, from 1:100 to 1:5. The biochip was prepared by immobilizing either S-wt or N-wt in the two signal regions and a-FITC in the negative control region.*

We fitted the initial part of the sensograms during the sample injection with a linear dependency for all curves in Figure 4 and retrieved the values of the slopes. Figure 5 shows the plots of the values of the fitted slopes versus the serum dilution. A linear dependency of the slope is found, which is in agreement with the kinetics predicted for a bivalent reaction of the kind[42,43]:

$$A + L \underset{k_{d1}}{\overset{k_{a1}}{\rightleftharpoons}} AL \qquad AL + L \underset{k_{d2}}{\overset{k_{a2}}{\rightleftharpoons}} AL_2 \qquad (1)$$

where A is the target antibody in the serum and L is the probe protein at the biochip surface. The bivalent nature of the binding kinetics, for large densities of the immobilized probes, was assessed with SPR assays reported in the Supplementary Material. The coupled equilibrium reactions give rise to the following differential equations for the binding kinetics:

$$\begin{cases} \frac{d[L]}{dt} = -2k_{a1}[A][L] + k_{d1}[AL] - k_{a2}[AL][L] + 2k_{d2}[AL_2] \\ \frac{d[AL]}{dt} = 2k_{a1}[A][L] - k_{d1}[AL] - k_{a2}[AL][L] + 2k_{d2}[AL_2] \\ \frac{d[AL_2]}{dt} = k_{a2}[AL][L] - 2k_{d2}[AL_2] \end{cases} \quad (2)$$

where [A], [L], [AL] and [AL$_2$] are the concentrations of A, L and the bound complexes A+L and A+2L, $k_{a1}$ and $k_{a2}$ (M$^{-1}$s$^{-1}$) and $k_{d1}$ and $k_{d2}$ (s$^{-1}$) are association and dissociation constants, respectively. When summing the last two equations in (2) we get:

$$\frac{d([AL]+[AL_2])}{dt} = 2k_{a1}[A][L] - k_{d1}[AL] \quad (3)$$

During the initial growth of the kinetic curves $[AL] \sim 0$ and $[L] = [L_{max}]$:

$$\left.\frac{d([AL]+[AL_2])}{dt}\right|_{t=0} = 2k_{a1}[A][L_{max}] \quad (4)$$

Equation (4) can be expressed in terms of the quantities measured in the assays:

$$slope = \frac{d\Delta\theta}{dt} \propto 2k_{a1} C\, 2\Delta\theta_{max} \propto 4k_{a1}\, C_0\, dilution\, \Delta\theta_{max} \propto dilution \quad (5)$$

where $C_0$ (M) is the concentration of the antibodies in the crude serum, $\Delta\theta \propto [AL] + [AL_2]$ is the BSW resonance shift (in pixel), $\Delta\theta_{max} \propto [L_{max}]/2$ is the maximum resonance shift when the probes at the surface are bivalently saturated by the antibodies. Equation (5) predicts the linear dependency of the initial slope of the sensograms with respect to the dilution, as found experimentally in Figure 6. This result confirms what is observed in Figure 5 but, on the other hand, from the linear fits it is not possible to infer any information on the association constant $k_{a1}$ since the proportionality coefficient is given by $k_{a1}\, C_0\, \Delta\theta_{max}$ where all three parameters are unknown.

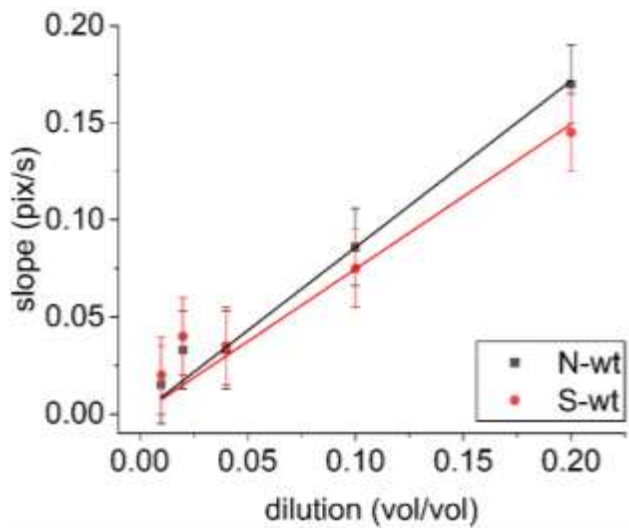

*Figure 5. Plot of the slope of the linear fit of the initial part of the differential sensograms shown in Figure 4 as a function of the serum dilution, for both the N-wt and S-wt regions. Lines represent the linear best fits of the data.*

**Comparative assays with BSW and MRR and SPR biochips**

In Figure *6*(a) and (Figure *6*b), we show the comparison sensograms recorded in assays carried out with either a BSW or MRR biochips in triplicate, respectively. The results were obtained for biochips that were chemically functionalized by the very same procedures with the same reactor and on whose active surfaces we immobilized two different antigens (S-wt, N-wt) and one negative reference (FITC), in separate BSW regions and MMR, respectively. Sample S404 was used for both assay formats. In the BSW case the biochip was first primed with AWB and then received a sample injection at 12.5 μl min$^{-1}$. In the MRR case the biochip was initially empty; after allowing a pre-wash solution to flow, the sample was added to the nanopillar microfluidic card in which flow occurs via capillary action, without control on the injection speed; the approximate flow rate was evaluated to be 5.2 ± 2.0 μl min$^{-1}$. The results of the assays are consistent between the two techniques and agree with the ZIVA values reported in the Table 1. The sensors discriminate the levels of S-wt and N-wt specific antibodies in the injected serum sample. In the case of the MRR biochips, we recorded a residual differential signal of about 100 pm that is due to nonspecific binding that isn't corrected by the negative control.

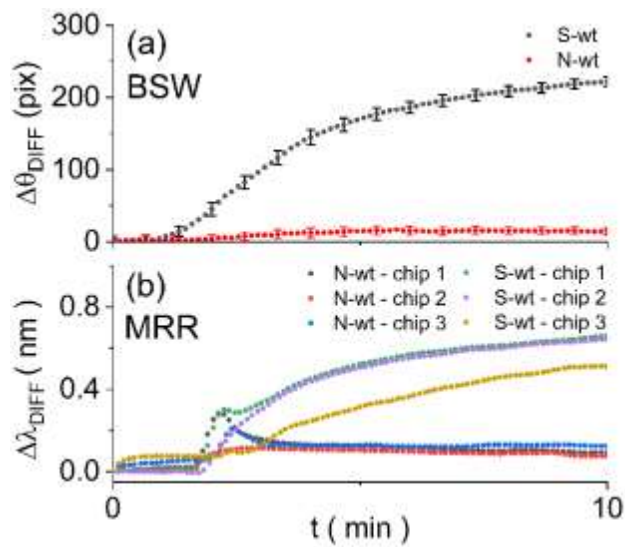

Figure 6. *Comparison of differential sensograms recorded for a 1:5 diluted S405 serum in BSW (a) and MRR (b) assays carried out with biochips bearing two signal regions (N-wt, S-wt) and a reference region (FITC). The MRR assays are in triplicate obtained with different biochips.*

Similarly to Figure *6*, Figure 7 displays the sensograms recorded during assays using (a) BSW, (b) MRR, or (c) SPR biochips. These results were obtained by injecting a synthetic sample containing the monoclonal antibody a-BA.5 Spiked in AWB at 10 µg ml$^{-1}$. The injection flow rates were comparable and are detailed in Table 2. For the BSW (a) and MRR (b) cases, two different antigens (S-BA.5 and S-O) and one negative reference (FITC) were immobilized on the active surfaces in distinct regions. In contrast, for the SPR (c) case, the S-BA.5 antigen was immobilized in the signal region, while the reference region was blocked with 1 M ethanolamine in water. Notably, the surface chemical functionalization in the SPR case differs from the others, as it is dextran-based and utilizes a 3D capture volume.

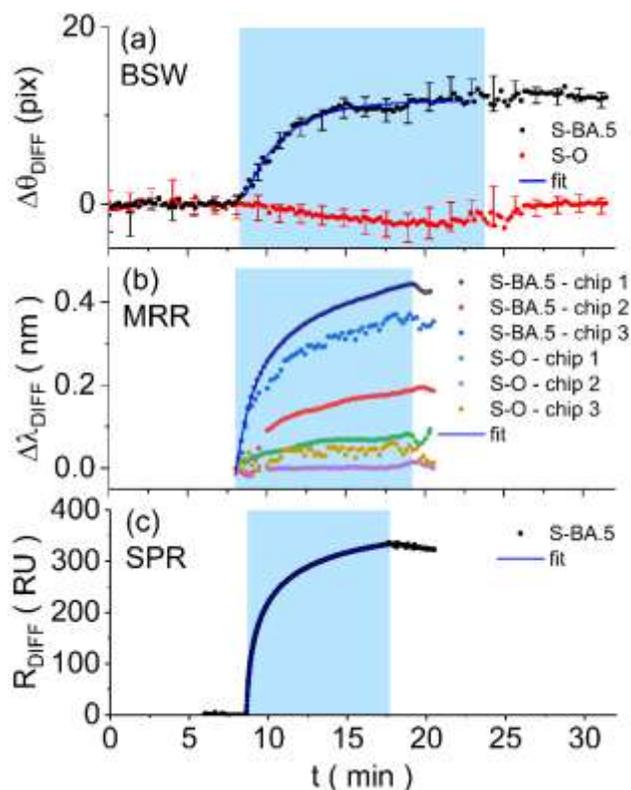

*Figure 7. Comparison of differential sensograms recorded for a 10 μg ml$^{-1}$ solution of a-BA.5 in AWB in BSW (a), MRR (b) and SPR (c) assays. In the (a) and (b) cases the biochips had two signal regions (S-BA.5, S-O) and a reference region (FITC). The MRR assays are in triplicate obtained with different biochips In the (c) case the SPR biochip had a signal (S-BA.5) and a reference (ethanolamine) region. The blue curves are the fits with either a single (a) or a double exponential (b,c).*

*Table 2. Time constants of the exponential fits of the association curves in Figure 7.*

|  | BSW | MRR | SPR |
|---|---|---|---|
| **Injection rate (μL min$^{-1}$)** | 12.5 | ~5 | 10.0 |
| **τ (min)** | 2.55±0.04 | 7.5 ± 0.7 (53%) | 3.01 ± 0.02 (75%) |
|  |  | 1.05 ± 0.04 (47%) | 0.35 ± 0.01 (25%) |

Despite their differences, the three techniques yielded comparable results and demonstrated rather consistent responses. The analysis of the exponential signal growth during the binding reaction is summarized in Table 2. Within experimental error, the sensograms were fitted using either a single exponential (a) or a double exponential (b, c) function; the resulting time constants and relative weights (in %) are provided in Table 2.

While the binding kinetics are influenced by the specific characteristics of each microfluidic system— precluding a strict direct comparison— the time constants for the primary components, those with the largest weights in the BSW and SPR cases and the mean in the MRR case that shows similar weights for the two components, remain comparable. Probably, the long time constant component in the MRR case is likely influenced by the absence of active flow control in the MRR microfluidic card.

**CONCLUSIONS**

In conclusion, the present study successfully demonstrates a comparative performance analysis of Bloch surface wave (BSW) and microring resonator (MRR) biosensors for the label-free detection of SARS-CoV-2 antibodies. By establishing a new BSW readout apparatus in close proximity to an existing MRR platform, the research conducted assays under identical conditions to ensure the highest degree of comparability. Both the BSW and MRR platforms demonstrated the capability for rapid, direct, and quantitative detection of anti-Spike and anti-Nucleocapsid antibodies in human serum without requiring secondary labels. The results obtained from both platforms were consistent with each other and showed strong agreement with existing ZIVA scores for the tested serum samples. Furthermore, the kinetics of the binding reactions (time constants) for BSW, MRR and SPR were found to be approximately in the same range. The BSW biochips exhibited excellent batch-to-batch repeatability, as evidenced by nearly identical responses when testing different chips with the same serum samples on different days. Investigations into the reusability of BSW biochips revealed that while surfaces functionalized with the Nucleocapsid

RBD probes could be successfully regenerated for multiple uses, those using the Spike RBD probes could not be effectively re-used due to the high affinity of the antibodies, leading to surface saturation. The initial binding kinetics of the BSW sensors showed a linear dependency on serum dilution, confirming the predictability of the reaction kinetics for both Nucleocapsid RBD and Spike RBD regions.

In summary, both BSW and MRR technologies offer sensitive and efficient analytical solutions for serological surveillance. Their ability to be integrated into low-cost, disposable formats positions them as highly promising next-generation tools for clinical diagnostics and the monitoring of immune responses.

## ASSOCIATED CONTENT

The following files are available free of charge.

BSW_Rochester_Supplementary_V10.docx (Supplementary Information, docx)

Covid_ROC_DataAnalysis_Chip_12_1.mp4 (movie, mp4)

## AUTHOR CONTRIBUTIONS

The manuscript was written through contributions of all authors. All authors have given approval to the final version of the manuscript.

## FUNDING SOURCES

This research was funded by the Italian Ministry of Research, under the complementary actions to the NRRP "D34health - Digital Driven Diagnostics, prognostics and therapeutics for sustainable Health care" Grant (# PNC0000001).(project code B53C22006120001) and by the US National Institute of Standards and Technology (NIST) Rapid Assistance for Coronavirus Economic

Response (RACER) program, grant number 70NANB22H015, as funded under the American Rescue Plan.

ACKNOWLEDGMENTS

The authors acknowledge technical and friendly discussions with Daniele Chiappetta, Paola Di Matteo, Michael Bryan, Alanna Klose, John Cognetti, Daniel Steiner, Joseph Bucukovski.

# Comparative performance of three optical biosensing platforms for SARS-CoV-2 antibodies detection in human serum

*Agostino Occhicone[1,2,‡], Alberto Sinibaldi[1,2,‡,*], Peter Munzert[3], Jordan N. Butt[4], Ethan P. Luta[6], Diego M. Arévalo[4], and Francesco Michelotti[1], and Benjamin L. Miller[5,6]*

1. SAPIENZA Università di Roma, Dipartimento di Scienze di Base e Applicate per l'Ingegneria, Via A. Scarpa 16 00161, Roma, Italy
2. Italian Institute of Technology, Centre for Life Nano and Neuro Science, Viale Regina Margherita 291, 00161, Roma, Italy
3. Fraunhofer Institute for Applied Optics and Precision Engineering IOF, Albert-Einstein-Str. 7, Jena 07745, Germany
4. Department of Chemistry, University of Rochester, Rochester, NY 14627, USA
5. Department of Dermatology, University of Rochester Medical Center, Rochester, NY 14642, USA
6. Materials Science Program, University of Rochester, Rochester, NY 14642, USA

*Supplementary Information*

* Author to whom correspondence should be addressed to francesco.michelotti@uniroma1.it

**Characteristics of the BSW biochips**

The BSW biochips used in the assays reported in the main manuscript are constituted of a glass substrate and a 1DPC deposited on top of it. They are designed to operate when the 1DPC surface is in contact with an aqueous medium (n=1.332). They bear the characteristics that are described in Figure SI - 1.

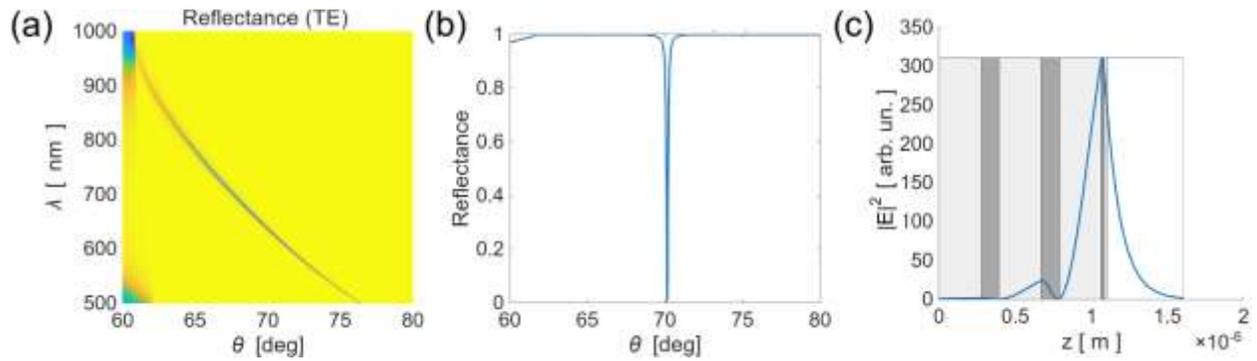

*Figure SI - 1. (a) Calculated TE reflectance for the design 1DPC of the BSW biochips. (b) Angular dependency of the TE reflectance calculated at $\lambda$ = 632.8 nm with the BSW resonance. (c) Calculated distribution of the squared field amplitude $|E|^2$ across the 1DPC under resonant excitation conditions. The field is peaked at the interface between the 1DPC and the external aqueous medium.*

Figure SI - 1a shows the calculated reflectance for TE polarized light impinging on the 1DPC from the substrate side. The reflectance is mapped as a function of $\theta$, the incidence angle in the substrate, and $\lambda$ the vacuum wavelength of the exciting beam. The ($\lambda$,$\theta$) dispersion of the BSW appears as a dark curve that starts from the TIR edge of the substrate/water interface at about $\lambda$=1000 nm and extends down to about 76 deg at $\lambda$ = 500 nm. The calculations were performed by means of a custom Matlab based transfer matrix method[1] code. Figure SI - 1b is the plot of the reflectance at the operation wavelength of the biosensing platform $\lambda$ = 632.8 nm. The calculated resonance width is very narrow and is determined by the optical losses of the materials constituting the 1DPC. The resonance width that is usually found experimentally is broader, mainly due to scattering losses arising from physical imperfections of the 1DPC and from the large inhomogeneity of the protein layers eventually immobilized at the 1DPC surface. Finally, Figure SI - 1c shows the squared electric field (TE) amplitude $|E|^2$ across the 1DPC when the biochip is excited at $\lambda$ = 632.8 nm and at the BSW resonant excitation angle. The plot is superimposed over the geometry of the 1DPC. Under resonant conditions $|E|^2$ is 300 times the value associated to the incoming field. The evanescent tail penetrating inside the external aqueous medium, on the right hand side of the plot, is used for label-free sensing purposes. The penetration depth is about 120 nm.

**BSW biochips readout platform.**

A sketch of the optical apparatus used to operate the BSW biochips is shown in Figure SI - 2. The beam emitted at $\lambda = 632.8\ nm$ by a low power cw He-Ne is polarized along the TE direction and expanded by means of a Galilean telescope (53x microscope objective (MO) and $f_0$ = 100 mm lens). In the focal plane of the telescope a rotating scattering disk is used to destroy the spatial coherence of the laser beam; the rotation speed was adjusted so as the CMOS integration time gives rise to speckle-free images. A rectangular diaphragm is used to select a quite homogeneous portion of the beam. A cylindrical lens $f_1$ = 100 mm is used to focus the illumination laser beam onto the coupling prism within an angular range $\Delta\theta \sim 6°$. The reflected beam is collected by a second cylindrical lens $f_2$ = 150 mm, which is acting as a Fourier lens; therefore, each pixel of the detector rows, lying in the incidence plane, corresponds to an angular component of the reflected beam. A third cylindrical lens $f_3$ = 70 mm, oriented perpendicular to $f_1$ and $f_2$, images the sensor surface on the array detector; therefore, each pixel of the array columns corresponds to a position along the illuminated region. The detector was an 8 bit CMOS array detector (Thorlabs DCC1645C). The optical configuration sets a width of the angular detection range of 1 deg along the largest dimension of the CMOS array (1280 pixel), therefore 1 pixel corresponds to an angular width of 0.775 mdeg. Along the other axis of the CMOS array (1024 pixel) the position along the illumination strip is obtained, where the imaged region is about 6.8 mm wide; binning 32 was applied along such a direction, giving rise to 32 spots along the strip.

The management of biological fluids and their injection into the microfluidic channel of the biochips made use of a Fluigent system.

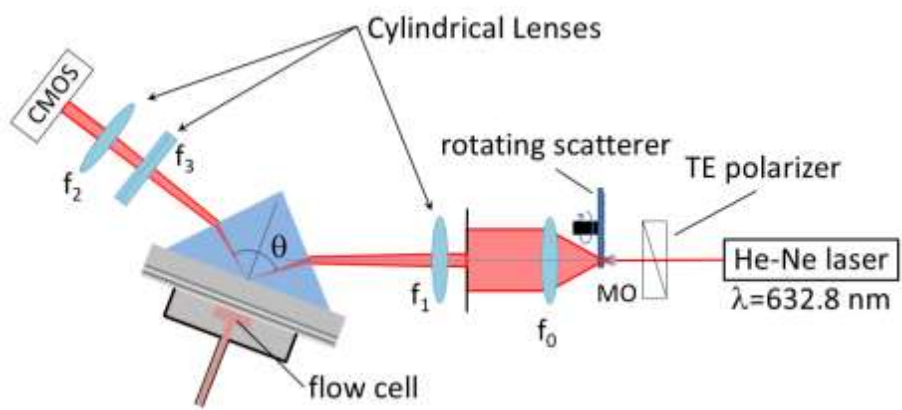

*Figure SI - 2. Sketch of the optical apparatus setup at the University of Rochester Medical Center to operate the BSW biochips.*

Figure SI - 3a shows the CMOS image acquired in AWB buffer at the beginning of an assay that was carried out with a BSW biochip where we immobilized N-wt, S-wt and a-FITC proteins in three different regions along the sensitive area. Such a biochip was used in the assay shown in Figure 4 in the main article. The three regions are marked with colored lines. The excitation of a BSW is witnessed by the black strips appearing in the image, where the reflectance is decreased. In each region the BSW resonance is in a different angular position, measured in camera pixel, due to inhomogeneities arising from the chemical functionalization, protein immobilization and BSA blocking procedures. Figure SI - 3b shows the reflectance profiles obtained by averaging intensity of several spots contained between the colored lines in the CMOS image. The color of the curves corresponds to the color of the sensitive region. The error bars are the standard deviation of the mean of the spots. The angular range detected by the CMOS camera is 1 deg, showing that the BSW resonance is indeed very narrow compared, for example, the resonance width observed in SPR platforms, which is about 1.7 deg.[2] Compared to the numerical simulation shown in Figure SI - 1b, the resonances are broadened, mainly due to scattering effects. The depth of the resonant dips leads to almost complete extinction of the reflectance at resonance, owing to the optimized design of the 1DPC used to fabricate the BSW biochips.

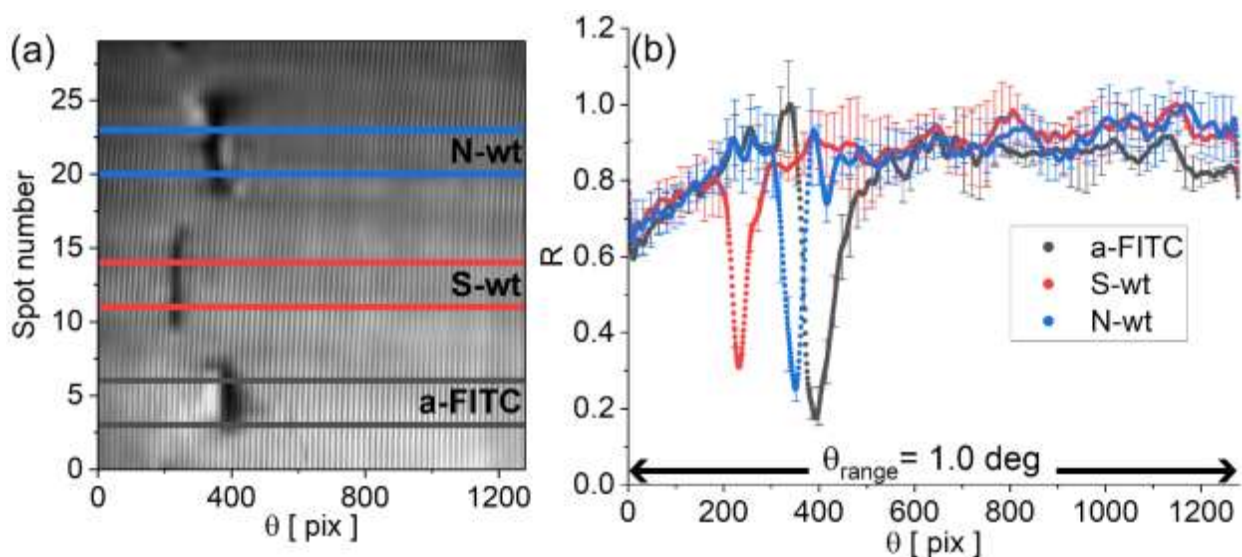

*Figure SI - 3. (a) CMOS image acquired at the beginning of an assay in AWB buffer. The colored lines delimit the sensing regions with N-wt, S-wt and a-FITC. (b) Reflectance curves obtained by averaging the spots inside each sensing region, with the same color coding.*

During a standard assay, the angular position of the minima of Figure SI - 3b are tracked in real-time to investigate binding of serum antibodies at the sensitive regions. In the movie available as Supplementary Information we show the time dependence of the CMOS image when a 1:5 diluted 405 serum is injected in the microfluidic channel of the biochip. In the video the real-time sensograms of the BSW resonance shifts in the three sensitive regions are shown too, demonstrating the operation of the read-out platform. The movie corresponds to the assay shown in Figure 2(c) in the main article.

**MRR readout platform.**

Instrument and data acquisition: The instrument used for acquisition of photonic sensor data, as well as the disposable micropillar microfluidic card, have been described previously.[3] In brief (Figure SI-4), light from a tunable infrared laser source (Keysight 81606A) is routed via single-mode fiber through a polarization controller (Thorlabs FPC561 with SMF-28 FC/PC connectors) to produce TE-polarized light, and from there to a custom diamond-turned optical element that sends the light vertically into focused grating couplers on the PIC. Output light from the PIC gratings is captured by a multimode fiber bundle (IDIL Optics; Thorlabs FP200ERT fibers) which connect to an optical power meter (Keysight N7745A). A combined vis/IR microscope stack is employed for PIC alignment. A computer is used for system control (Keysight Photonic Application Suite) and data acquisition.

Micropillar microfluidic card: Details of the injection-molded micropillar microfluidic card have been previously published. In brief, and with reference to the image shown in Figure SI-5, a 1 x 4 mm ring resonator PIC is integrated with a plastic card consisting of a sample zone for sample introduction, and wicking zone and wicking pad that acts as a sink for fluid that has passed by the photonic sensor chip. These two areas are connected by a channel with micropillars (60 microns in diameter, separated by 110 microns and at a 60º angle to each other) to regulate the flow of fluid under the PIC.

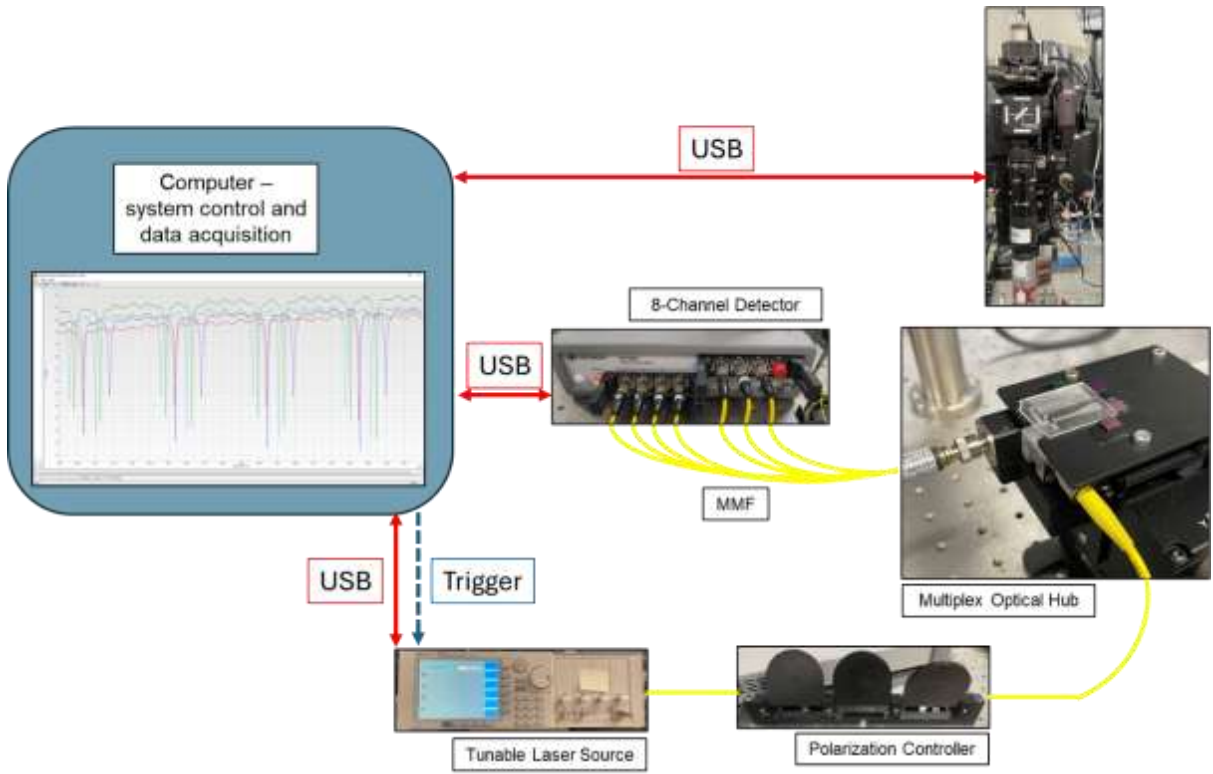

*Figure SI – 4. Schematic of the MRR optical system.*

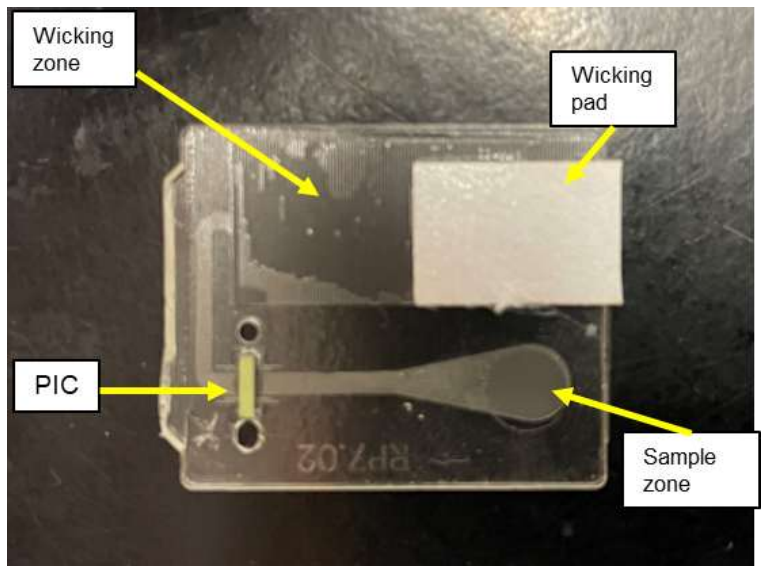

*Figure SI-5. Micropillar microfluidic card for passive delivery of fluid to the sensor PIC.*

**SPR characterization of the affinity of monoclonal antibodies a-BA.5 against S-BA.5.**

SPR assays for the determination of the affinity of a-BA.5 monoclonal antibodies against the S-BA.5 **RBD** were carried out with a Cytiva Biacore X100 SPR instrument and ready-to-use Cytiva sensor chips of the CM5 type, which featured a standard dextran surface chemical functionalization. The S-BA.5 RBD were immobilized using the 0.4 M EDC (1-ethyl-3-(3-dimethylaminopropyl)carbodiimide) and 0.1 M NHS (N-hydroxysuccinimide) procedure. The reference spot flow channel was blocked with 1 M ethanolamine. The immobilization concentration of the S-BA.5 RBD in sodium acetate buffer at pH 5 was 50 µg/ml, which gave rise to a RU differential signal of ~2500 with the CM5 sensor chip.

In Figure SI-6, we show the SPR response upon serial injection of the a-BA.5 diluted in HEPES runner buffer ( 10 mM HEPES free acid, 150 mM NaCl, 3 mM EDTA, 0.005% tween 20, pH 7.4) at 10, 3.33, 1.11, 0.370, 0.123 µg/ml. Figure SI-6(a) shows global data fitting assuming a 1:1 binding kinetics between the antibody antigen pair, showing that the 1:1 model does not provide a good quality fit and that the value found for the dissociation constant $K_D$ found (in the pM range) is beyond the limits of the SPR instrument. Figure SI-6(b) shows global fitting when assuming a bivalent binding model, which is plausible since the antibody can effectively bind to two different antigens at once, therefore increasing avidity.[4] This also explains the lack of a dissociation phase in the plots. The plot for bivalent fit is significantly better than the 1:1 case. The Cytiva fitting routine based on the bivalent binding model described in the main article, we found: $k_{a1} = 2.85 \cdot 10^5 \, M^{-1}s^{-1}$, $k_{d1} = 1.91 \cdot 10^{-5} \, s^{-1}$, $k_{a2} = 3.27 \cdot 10^{-5} \, RU^{-1}s^{-1}$, $k_{d2} = 2.28 \cdot 10^{-3} \, s^{-1}$, $R_{max} = 758.7 \, RU$. Since the interaction at the second site does not contribute to the SPR response, $k_{a2}$ is given in units of RU$^{-1}$s$^{-1}$, and could only be obtained in M$^{-1}$s$^{-1}$ if a reliable conversion factor between RU and M would be available. Similarly, a value for the overall affinity or avidity constant cannot be inferred. From the steady state signal plot as a function of concentration given in Figure SI-6(c) one can infer a value for the dissociation constant $K_D = 6.15 \cdot 10^{-9}\ M$.

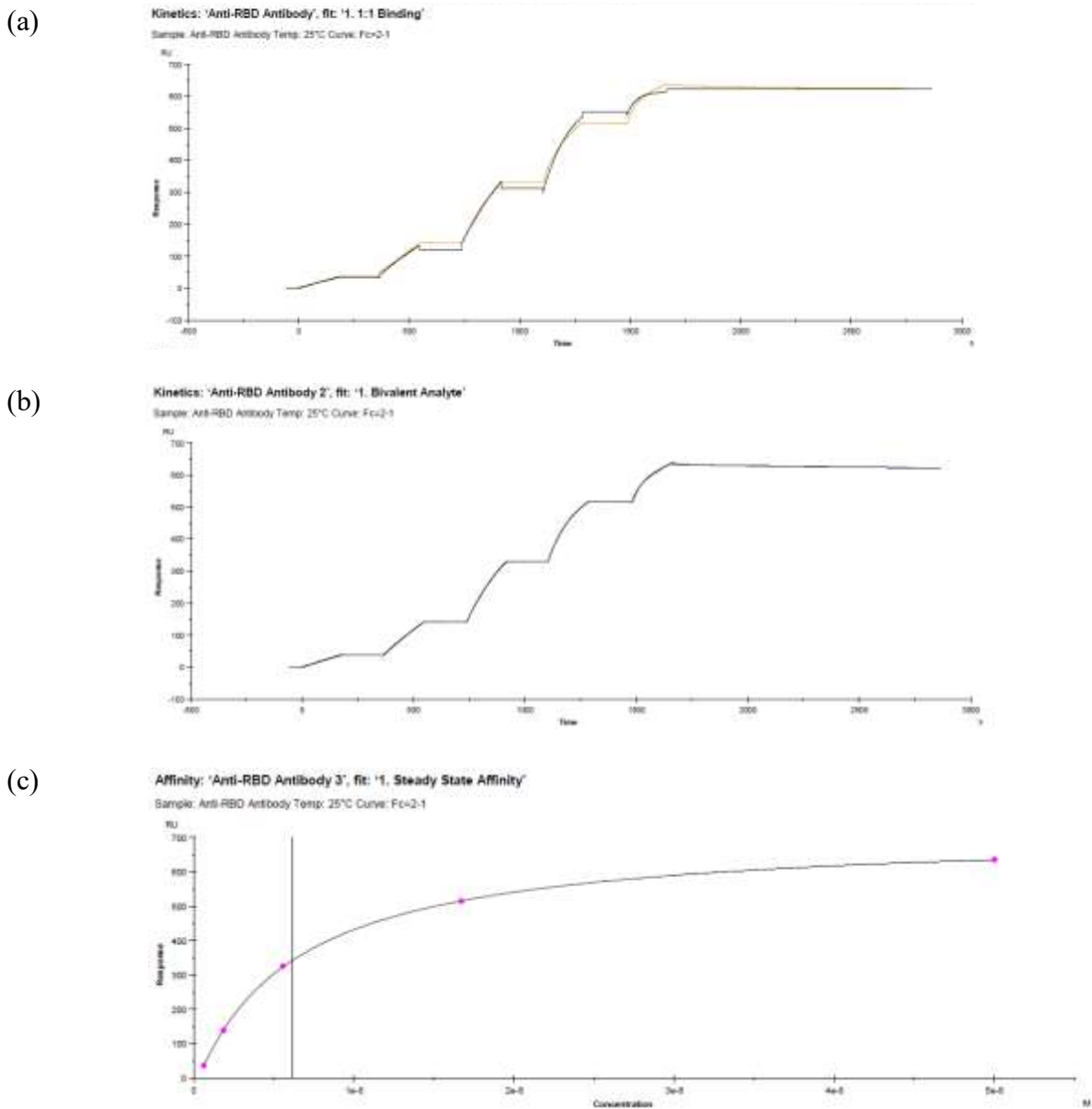

*Figure SI-6. (a-b) Sensogram recorded during a-S-BA.5/S-BA.5 RBD interaction when injecting different samples in a sequence at the following increasing concentrations of a-BA.5 in HEPES Buffer, 10, 3.33, 1.11, 0.370, 0.123 µg/ml. (a) monovalent 1:1 global fit, (b) bivalent binding model. (c) Steady state signal plot as a function of concentration.*